# Label-free video-rate micro-endoscopy through flexible fibers via Fiber Bundle Distal Holography (FiDHo)

Noam Badt[1] and Ori Katz[1,*]

[1] Department of Applied Physics, Hebrew University of Jerusalem, Jerusalem 9190401, Israel.
[*]orik@mail.huji.ac.il

**ABSTRACT**
Fiber-based micro-endoscopes are a critically important tool for minimally-invasive deep-tissue imaging. However, the state-of-the-art micro-endoscopes cannot perform three-dimensional imaging through dynamically-bent fibers without the use of bulky optical elements such as lenses and scanners at the distal end, increasing the footprint and tissue-damage. While great efforts have been invested in developing approaches that avoid distal bulky optical elements, the fundamental barrier of dynamic optical wavefront-distortions in propagation through flexible fibers, limits current approaches to nearly-static or non-flexible fibers. Here, we present an approach that allows holographic 3D bend-insensitive, coherence-gated, micro-endoscopic imaging, using commercially available multi-core fibers (MCFs). We achieve this by adding a miniature partially-reflecting mirror to the distal fiber-tip, allowing us to perform low-coherence full-field phase-shifting holography. We demonstrate widefield diffraction-limited reflection imaging of amplitude and phase targets through dynamically bent fibers at video-rates. Our approach holds great potential for label-free investigations of dynamic samples.

# INTRODUCTION

Flexible optical micro-endoscopes are an important tool for a wide variety of applications, from clinical procedures to biomedical investigations, where micron-scale structures such as single neurons are imaged [1, 2] or optically excited [3] at depths beyond the reach of conventional microscopes. Developing a flexible, 3D, video-rate, label-free micro-endoscope with a minimal footprint is thus a sought-after goal for minimally invasive deep-tissue imaging.

In the quest for this goal various small diameter micro-endoscopes have been developed in recent years. One set of solutions consists of micro-endoscopes that are based on single-mode fibers. while bend-insensitive, these require distal optical elements such as scanners and lenses [1], or spectral dispersers [4, 5] to produce an image. Such distal elements significantly enlarge the endoscope diameter, increasing tissue damage, and consequently limiting its use for minimally-invasive deep-tissue imaging.

Another set of solutions utilize imaging fiber bundles, also known as multi-core fibers (MCFs). These consist of thousands of cores closely packed together. Conventionally, each core functions as a single pixel, either by direct contact with the target or by the addition of a GRIN lens at the distal end [1, 2]. While straightforward to use, conventional bundle-based endoscopes suffer from limited resolution, pixelation and a small and fixed working distance with poor axial sectioning capability. Even though axial-sectioning can be improved by using temporal coherence gating [6, 7] or confocal or other structured illumination [8], many of the inherent limitations remain. These include pixelation artefacts, limited resolution, and the fixed working distance. Moreover, strong spurious reflections from the fiber tips limit the applicability of such approaches for label-free imaging.

An emerging set of approaches for lensless endoscopes that go beyond many of the conventional limitations are based on the measurement of the fiber transmission-matrix (TM), with or without an additional distal mask or diffuser, and the compensation of the complex coherent or incoherent transfer-function of these multi-mode systems, either digitally [10, 11, 12, 13, 14, 15] or using wavefront-shaping [16, 17, 18, 19, 20, 21, 22, 23]. However, speckle based incoherent imaging approaches suffer from low contrast of the acquired images [11, 12, 13, 14], and the practicality of solutions that are based on prior measurement of the TM without the addition of a distal element is limited to non-flexible fibers, since fiber bending leads to dynamic random phase distortions.

The TM may be estimated using a large number of measurements and complex computation, but thus do not allow video-rate imaging or imaging through dynamically bent fibers [24, 25, 26]. MCFs with bend-insensitive inter-core phase relations have been fabricated [27]. However, these rely on a relatively small number of single-mode cores, suffering from low collection-efficiency and fill-factor.

Related to these TM-based approaches, Czarske et al. [28, 29, 30]have demonstrated in-situ measurement of the fiber phase-distortions by adding a partially-reflecting mirror to the MCF distal-facet, and correction using a spatial light modulator (SLM). However, as imaging is performed by raster scanning a phase-corrected focus, correction of dynamic distortions is challenging, and fluorescence labeling is required. Holographic approaches to endoscopy that rely on external illumination and use the MCFs as intensity-only image-guides have been put forward over the years [31, 32, 33]. These require additional fibers for illumination, suffer from poor axial-sectioning, and limited FoV or coherent background due to twin image formation.

Here we present a simple approach that allows holographic 3D, calibration-free, bend-insensitive, coherence-gated, micro-endoscopic imaging utilizing commercially available MCFs. We achieve this by adding a miniature partially reflecting mirror to the distal fiber-tip (Fig. 1), which allows us to perform low-coherence full-field phase-shifting holography at video-rate, with diffraction-limited resolution, where the coherent illumination is provide by one of the MCF cores.

## RESULTS
**Principle**

The optical setup for realizing our approach is schematically depicted in Fig. 1. It is based on the excitation of a single MCF core by two delayed replicas of a low coherence illuminating beam (Fig. 1A). The two beams co-propagate in the same core along the MCF. At the MCF distal end, the reflection of the first arriving beam from the target interferes with the reflection of the second ("reference") beam from the distal partially reflecting mirror (Fig. 1B). The interference intensity pattern is collected by the MCF cores (Fig. 1C) and relayed back to the proximal end where it is imaged on a camera (Fig. 1A).

The object complex field at the distal facet is reconstructed from $N \geq 3$ camera images by low-coherence phase-shifting holography, which is performed by controlling the phase of the reference beam (Fig. 1A). The use of a low-coherence source allows effective rejection of all spurious reflections by setting the time delay between two excitation beams, $\tau$, to match the object distance from the mirror, $z_0$: $\tau = 2z_o/c$. Both the time delay $\tau$ and the phase-shifting are performed using the same interferometer that produces the two excitation beams.

The recorded hologram at the n-th phase, $I_n$, is the interference of the diffracted field from the object, $E_o$, and a known, fixed spherical reference field, $E_r$, that is reflected from the distal mirror: $I_n(x, y) = |E_o(x, y) + E_r(x, y)e^{i\varphi_n}|^2$, where $\varphi_n = \frac{n}{N} 2\pi$ is the phase added to the reference field. Importantly, the recorded interference intensity pattern is insensitive to fiber bending since the MCF faithfully relays the intensity images. Finally, the 3D object field is reconstructed from the field at the distal facet via back-propagation, as explained below.

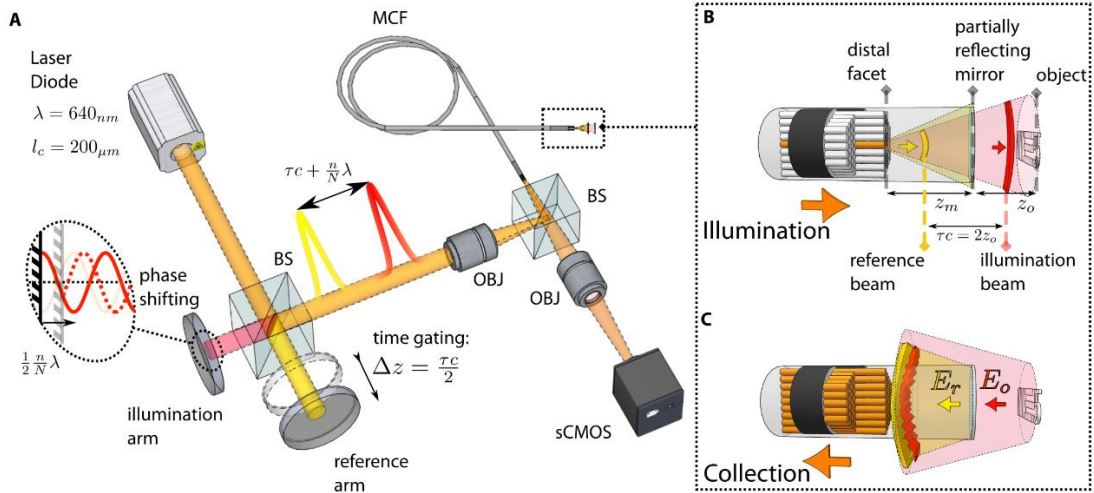

**Fig. 1. Setup and principle of distal holographic endoscopy.** (**A**) A short coherence laser (orange) is split into two delayed replicas: an illumination beam and a reference beam (red and yellow, respectively) that are coupled into a single core of a multicore fiber (MCF). (**B**) At the distal MCF facet, the illumination beam and the reference beam are reflected from the target object and a partially reflecting distal mirror, respectively. (**C**) The intensity pattern of the interference between the reflected beams is relayed by the MCF to the proximal side, where it is imaged on a camera (**A**). Due to coherence gating, setting the relative time delay between the two arms to match the object-mirror distance, $\tau = 2Z_o/c$, results in measured interference only between the reflected illumination from the object and the reflected reference from the mirror. The object complex field is retrieved from $N$ intensity-only proximal images, by phase-shifting the illumination arm (**A**, inset). See Fig. S1 for the full setup.

## Reconstruction process

The reflected object field at the fiber distal tip, $E_o$, is reconstructed from $N \geq 3$ camera frames, $I_n$ (A), by phase-shifting interferometry followed by division by the conjugate reference field, $E_r^*$ (B):

$$E_o(x,y) = \frac{1}{E_r^*(x,y)} \sum_{n=1}^{N} I_n(x,y) e^{i\varphi_n} \qquad (1)$$

where $E_r^*$ is approximated as a Gaussian beam from the theoretical first mode of a single fiber core.

3D reconstruction is then performed by back-propagation of the recorded field (D), $E_o$, to any desired distance from the fiber facet, $z_{prop}$, and normalization by the expected illumination field amplitude and phase $E_{illum}(x, y, z_{prop})$:

$$O(x, y, z_{prop}) = \frac{\mathcal{P}_{-z_{prop}}(E_o(x,y))}{E_{illum}(x, y, z_{prop})} \qquad (2)$$

Where $O$ is the object complex reflection coefficient, $\mathcal{P}_{-z_{prop}}$ is the operator for angular spectrum propagation by a distance $-z_{prop}$, and $E_{illum}$ is approximated as a Gaussian beam from the theoretical first mode of a single fiber core. A digitally refocused sharp image of the object amplitude and phase is obtained (Fig. 1D) at the propagation distance $z_{prop} = z_o + z_m$, where $z_o$ is the object-mirror distance and $z_m$ is the mirror-fiber distance (Fig. 1B). The object distance can also be found from the low coherence holograms by scanning the time delay, $\tau$, and plotting the total energy of the reconstructed field at each time delay (C).

As can be observed in (Fig. 3A-H), unlike conventional MCF-based endoscopy, the holographic reconstruction is un-pixelated, since the MCF pixelation occurs at a difference axial plane. The MCF pixelation at the fiber facet, which may introduce ghosting, is removed using simple interpolation (see "Digital filtering of MCF pixelation" section).

Beyond the 3D holographic amplitude and phase label-free reconstruction, FiDHo has several additional merits: it is insensitive to dynamic random phase distortions that are potentially introduced by fiber bending, the full field reconstruction requires only 3 frames, allowing video-rate imaging, and the low coherence gating improves axial-sectioning. In the next sections we experimentally demonstrate each of these merits.

## Phase shifting

$$I_n(x, y) = |E_o + E_r e^{i\varphi_n}|^2 \xrightarrow{1/E_r^* \sum_{n=1}^{N} I_n e^{i\varphi_n}} E_o(x, y)$$

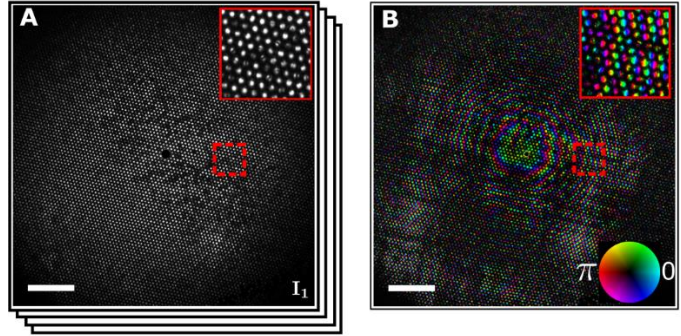

## Time gating

$$\int |E_o(x, y)|^2 \, dx dy$$

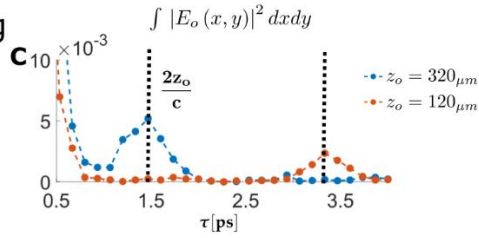

## Digital refocusing:

$$O = \frac{\mathcal{P}_{z-prop}(f(E_o(x,y)))}{E_{Illum}(x, y, z_{prop})}$$

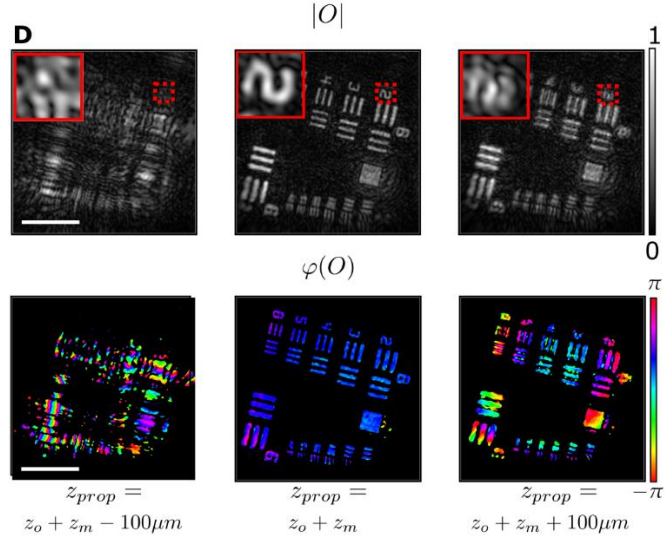

$z_{prop} = z_o + z_m - 100\mu m$     $z_{prop} = z_o + z_m$     $z_{prop} = z_o + z_m + 100\mu m$

**Fig. 2. Experimental holographic imaging of a resolution target.** (**A**) $N$ phase-shifted proximal intensity images $I_n(x, y)$ are used to retrieve the reflected object field on the distal facet, $E_o(x, y)$ (**B**). (**C**) Scanning the time delay between the reference and illumination arms reveals a peak of the total field energy at the correct target distance, as verified by placing a reflective target at two different distances (orange, blue); (**D**) Digital refocusing is achieved by back-propagating the measured distal field $E_o(x, y)$. The back-propagated field reveals the reflective USAF target in focus (top center) and with a flat phase (bottom center) at the correct distance ($z_{prop} = z_o + z_m$, $z_o = 340\mu m$), after normalizing by the illuminating field. At other propagation distances (left and right) the target is out-of-focus. (Insets, zoom-in on dashed rectangles). Scale bars: $100\mu m$

**Resolution and field-of-view**

To evaluate the system resolution and field of view (FoV), we performed several sets of experiments using reflective targets placed at different distances from the fiber (Fig. 3). Unlike conventional MCF-based microendoscopy, where each fiber core serves as one imaging pixel and the resolution is limited by the core-to-core pitch (Fig. 3A), in FiDHo the resolution is diffraction-limited and the images are unpixelated (Fig. 3C-D). Specifically, FiDHo easily resolves group 7 element 4 of a USAF resolution target, signifying resolution better than $2.7 \mu m$ (Fig. 3G). More than 2 times improvement over conventional contact mode (Fig. 3E). In addition to imaging the USAF resolution target at several distances (Fig. 3C,D), a precise quantification of the resolution and FoV were performed by imaging a knife-edge mirror and a large mirror respectively (Fig. 3I, J). Interestingly, and as expected from the theoretical analysis (see section S1), the resolution is diffraction limited by the fiber bundle numerical aperture (NA), and the FoV is half of the fiber diameter, independent of imaging distance, for distances of $z_{obj} < D_{fiber}/2NA - z_m$, where $D_{fiber}$ is the MCF diameter. For our experimental parameters ($z_m \approx 2mm$, $D_{fiber} \approx 600\mu m$), the measured resolution and FoV ($300\mu m$, width) indicate an effective $NA_{eff}$ of $\sim 0.15$, which is in line with the fiber parameters (Schott 153385) and the interpolation performed (see "Digital filtering MCF pixelation" section).

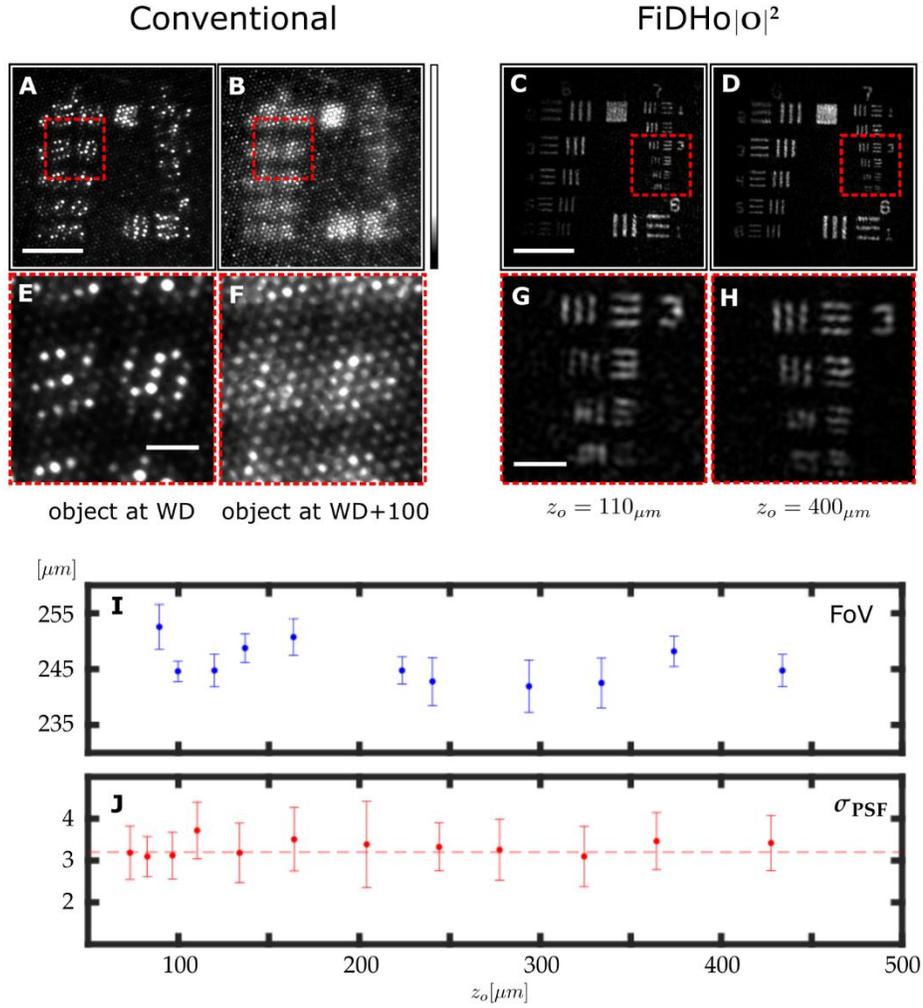

**Fig. 3. Resolution and field-of-view characterization.** Unlike conventional MCF-based microendoscopy, where each fiber core serves as one imaging pixel, the resolution is limited by the core-to-core pitch (**A**), and the object (a USAF 1951 target) is pixelated and in focus only at a specific working-distance (A, B). In FiDHo (**C**, **D**), the object is unpixellated and digitally focused on any distance, resolving features with a resolution that exceeds the fiber pixel pitch (**E-H**, zoom-in on dashed rectangles in A-D). (**I**) The system FoV, as measured by $1/e^2$ of the reconstructed intensity of a flat mirror (**J**) Imaging resolution as a function of depth, as retrieved from a knife-edge measurement (see section S1). Scale bars: (A-D) $100 \mu m$, (E-H) $25 \mu m$

**Phase-contrast imaging**

The holographic nature of FiDHo has inherently phase-contrast imaging capability. Phase-contrast imaging is important for the study of a wide variety of biological targets that presents very low reflection or absorption contrast and is not available in conventional endoscopes [34].

To demonstrate phase-contrast imaging, we imaged human cheek cells placed on a glass slide immersed in water. While the cells are not visible in the reconstructed field amplitude (Fig. 4A), the phase reconstruction clearly reveals the cells (Fig. 4B), as is validated by the transmission image of the sample recorded by a reference distal camera (Fig. 4C,D).

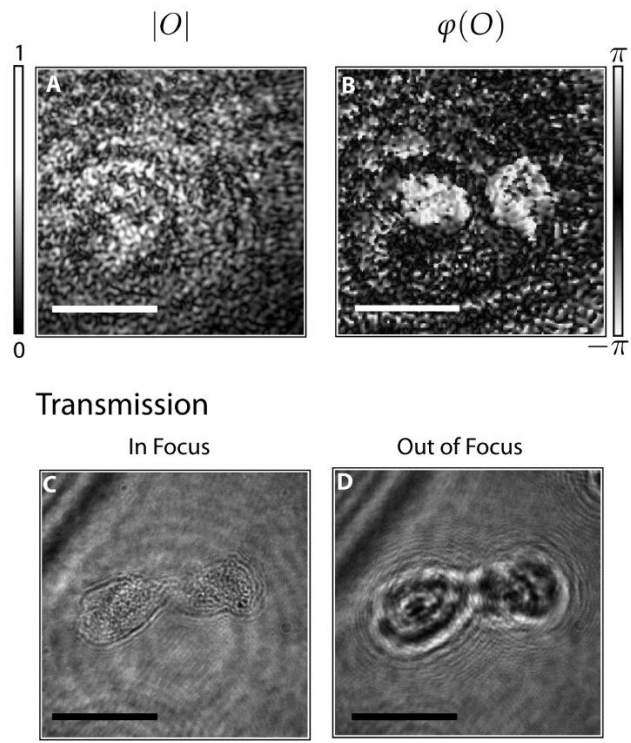

**Fig. 4. Phase-contrast imaging** (**A**, **B**) Reconstructed image of two cheek cells, placed in water on a microscope slide. The cells, which are not visible in the reconstructed field amplitude (**A**), are clearly visible in the reconstructed phase (**B**). (**C, D**) In-focus (**C**) and out-of-focus (**D**) widefield transmission microscope image of the cells. Scale bars- $80 \mu m$

.

**Video-rate and sensitivity to bending**

An important advantage of FiDHo is the high imaging speed, which results from the requirement for only three camera frames per reconstruction. The imaging speed is practically limited by a third of the maximal camera framerate. Supplementary Movie S1 demonstrates reconstruction of a moving target at 50 FPS. Some frames are shown in Fig. 5A.

An additional major advantage of FiDHo is the very low sensitivity to fiber bending, an important requirement for *in-vivo* and freely behaving animal studies [3, 35]: In the illumination path, since both the illumination and reference beams travel through the same mode of a single core, the illumination has inherently low sensitivity to bending. In the collection path, since only the intensity of the light is collected though the MCF, the detection is essentially insensitive to fiber bending. The very low sensitivity to bending together with the high imaging speed enables imaging through dynamically bent fibers, as is demonstrated in Supplementary Movie S2 and Fig. 5B.

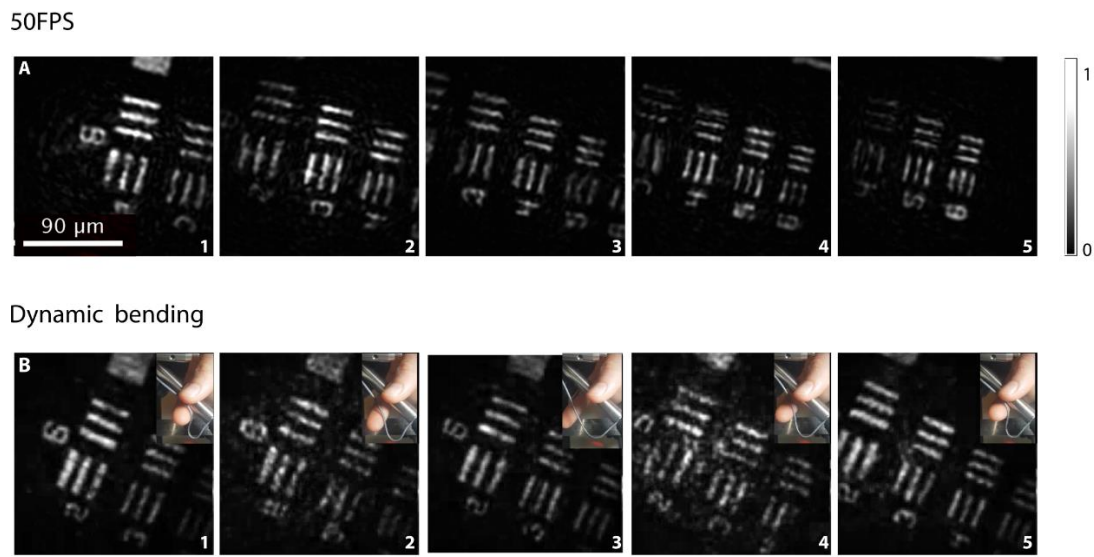

**Fig. 5. Dynamic imaging at video-rate.** (**A**) Selected frames from a real-time video at 50 frames-per-second (FPS) of a moving resolution target (see Supplementary Movie S1). (**B**) same as (**A**) when the fiber is dynamically bent, showing the insensitivity to fiber orientation, and low sensitivity to bending (see Supplementary Movie S2)

**3D Imaging**

The low-coherence time gating of FiDHo allows axial sectioning of 3D objects [36]. To demonstrate this, we have performed several experiments whose results are presented in Fig. 6. The experiments include imaging of a sample composed of two stacked USAF targets with an axial separation of $300 \mu m$ (Fig. 6A) and a thick chicken breast tissue (Fig. 6B). In both experiments, the low-coherence gating together with the holographic reconstruction allows axial-sectioning, depth measurement, and digitally refocused reconstruction of all sample planes.

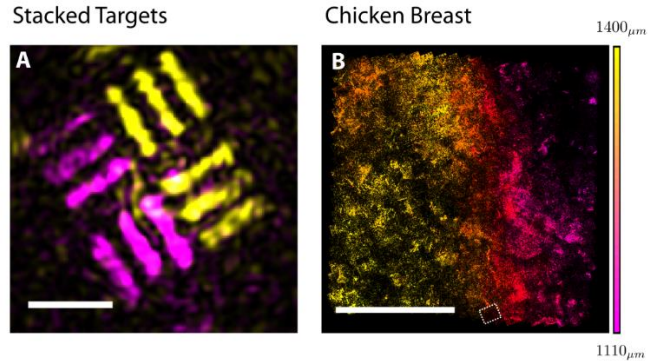

**Fig. 6. Imaging three-dimensional objects**(**A**) A 3D image of a target composed of two stacked resolution targets with a spacing of $\sim 300 \mu m$ reconstructed by super-posing two reconstructed images acquired with two appropriate time delays, $\tau = 2z_o/c$, with $z_o = 1030 \mu m$ (pink) and $z_o = 1320 \mu m$ (yellow). (**B**) An image of a chicken breast sample, reconstructed by stitching $11 \times 11$ sub-images, each with the fiber field of view (dashed rectangle). The varying depth of each sub-image is retrieved from the time-delay scan. Scale bars: A - $50 \mu m$ B - $500 \mu m$

## DISCUSSION

We have demonstrated an approach for 3D lensless endoscopy that possess a unique set of advantages, not jointly attainable with current approaches. These include video-rate 3D diffraction-limited, axially sectioned label-free imaging, in a calibration-free, bend-insensitive system, without any moving distal elements. Importantly, all these advantages are obtained in a simple system that employs commercially available MCFs, and with a straightforward non-iterative computational reconstruction.

For optimal imaging performance the parameters of the MCF, distal-mirror, and powers ratio of the object- and reference-arm should be set according to several considerations. These considerations are analyzed in detail in section S2. To summarize, the fiber diameter, $D$, dictates the FoV and maximal depth for diffraction-limited resolution (Fig. 3), and should be as large as allowed by the selected application. The $NA$ of the MCF cores dictates the system resolution. Bend-sensitivity is minimized by choosing MCF cores with low inter-core crosstalk, and single-mode propagation. The distal mirror distance, $z_m$, should be large enough such that the reference-wave phase is approximately constant over each individual core, yielding: $z_m \gg \frac{Dd}{2\lambda}$, where $d$ is the core diameter. Finally, the optimal mirror reflectively is dependent on the target reflectively and is approximately $2\% - 14\%$ for a reflected power fraction of $0.1\% - 4\%$ from the target.

A potential attractive application of FiDHo is for neuro-imaging, where a common practice is to image the mouse brain through a cannula incorporating a glass coverslip [37]. The cover-slip can be naturally utilized as the partially reflecting mirror, without any modifications to the distal probe.

The presented imaging results may be improved via angular-compounding using multiple illuminations from different cores, at the expense of framerate. Additional information that may extend the FoV and improve resolution can be extracted from the modes-distribution inside each core [38, 39], instead of the current integration of their energies. The use of high framerate and large well-depth cameras will improve the signal to noise and imaging speed, as in full-field OCT [40]. Finally, the image reconstruction resolution and fidelity can be improved by employing compressed-sensing reconstruction algorithms that incorporate prior knowledge on the target object structure [41].

## MATERIALS AND METHODS

**Experimental design**

Fig. 1. depicts a simplified sketch of the experimental setup, where the full setup design is presented in Fig. S1 and described below. The illumination is provided by a diode laser at a wavelength of $\lambda = 640_{nm}$ and coherence length of $l_c \sim 400_{\mu m}$ (iBeam-smart-640s, Toptica, $\sim 1_{mW}$ at proximal facet). The illumination beam is split by a Mach-Zehnder Interferometer (MZI) using a polarizing beam splitter (PBS1, Thorlabs PBS251), and a non-polarizing 50-50 beam splitter (BS2, Thorlabs BS013). The powers ratio in the two arms is controlled by a half wave-plate (HWP1, Thorlabs WPH10M-633) placed before PBS1, and the polarizations are re-aligned by a second half-wave plate (HWP2, same as HWP1). The two arms need to be perfectly aligned, such that the two delayed beam replicas couple to the fundamental mode of a single MCF core. This is assured by using two separate 4-f telescopes (L1, L2. total of four $f = 100_{mm}$ Thorlabs LA1509) placed in each of the MZI arms. Alternatively, the delayed replicas can be inherently aligned using a single-mode fiber (SMF) based MZI.

The two aligned beams exiting the MZI are coupled into a single core of the MCF using another telescope (L3, Thorlabs LB1901, $f = 75_{mm}$ and L4, Thorlabs LA1979, $f = 200_{mm}$) and a long working-distance objective (OBJ1, Mitutoyo 378-803-3, M Plan Apo 10x, 0.28 NA). A non-polarizing 50-50 beam-splitter (BS3, Thorlabs BP150) is placed between OBJ1 and the MCF proximal facet. The MCF proximal facet is imaged through BS3 onto an camera (sCMOS, Andor Zyla 4.2Plus) by another long working distance objective (OBJ2, same as OBJ1), a lens (L5, Thorlabs LA1979, $f = 200_{mm}$) and a telescope with tunable magnification (ZOOM, Navitar 12X Zoom Lens System).

In order to minimize the proximal facet reflections in the camera image, two orthogonal linear polarizers (LP1, LP2: Thorlabs LPVISE100) are placed on the two ports of BS3 to effectively perform cross-polarized detection.

The reference arm length in the MZI is controlled by a two-mirror delay-line using a fine piezo-motor translator (Thorlabs PIA25 actuator and TIM101 controller) mounted on a larger translator (Thorlabs Z825B actuator and KDC101 controller) for phase-shifting and time gating, respectively.

At the distal end, the small partially reflecting mirror surface is placed at a distance of $z_m = 2_{mm}$ from the fiber distal facet. The partial reflective mirror was fabricated by E-beam evaporation (EBPVD) of a $10_{nm}$-thick layer of Titanium on a $1_{mm}$ glass slide. An additional $1_{mm}$-thick glass slide was used as a spacer, and an optical immersion oil with $n = 1.52$ refractive index was placed between the glass spacer and the fiber and mirror to reduce unwanted reflections. The targets were placed at different distances behind the partially reflecting mirror, either by holding the samples in air or with immersion oil or water between the mirror and the samples. An additional distal camera (CMOS, Allied Vision Mako U-130B) is used for acquiring the ground-truth images of the sample using a microscope objective (OBJ3, Olympus UPlanFL n, 10x 0.3 NA) and a lens (L6, Thorlabs LB1676, $f = 100_{mm}$).

**Suppression of spurious reflections**

Since the system is designed to image weakly reflecting objects, it is very sensitive to unwanted and spurious reflections from the MCF proximal facet. The cross polarized detection effectively reduces most of the proximal reflective background. Thanks to random birefringence of the MCF cores [42], the reflected signal from the object is randomly polarized after propagation through the MCF, and thus the cross-polarized detection only halves the reflected object power on average.

Spurious reflections at the MCF distal facet contribute only to strong intensity in the single excited core, which is spatially filtered digitally.

Two additional sources of unwanted reflections that are naturally present but do not appear in the simple form of eq.1 are the reflection of the object arm from the distal mirror, and the reflection of the reference beam from the object. The interference terms arising from these two unwanted reflections are effectively suppressed by low-coherence time gating. In general, the interference on the fiber distal facet is the result of interference of four reflected waves: two desired waves and two undesired waves. The two desired reflected waves are the reflection of the reference beam from the distal mirror $E_{R,m}$, and the reflection of the object illumination beam from the object $E_{I,o}$. The two undesired reflected waves are the reflection of the reference beam from the object $E_{R,o}$, and the reflection of the object illumination beam from the distal mirror $E_{I,m}$. Thus, the intensity interference pattern at the fiber distal (and proximal) facet is:

$$I = \left|E_{R,o}e^{i\varphi_n} + E_{R,m}e^{i\varphi_n} + E_{I,o} + E_{I,m}\right|^2 \quad (3)$$

where $\varphi_n$ is the phase-shifting phase.

As a result, the phase-shifting reconstructed hologram will be composed of 4 interference terms:

$$\underbrace{E_{I,o}E_{R,m}^*}_{A} + \underbrace{E_{I,m}E_{R,m}^*}_{B} + \underbrace{E_{I,o}E_{R,o}^*}_{C} + \underbrace{E_{I,m}E_{R,o}^*}_{D} \quad (4)$$

The first term (A) is the desired signal, composed of the interference of the object illumination beam reflected from the object with the reference beam reflected from the mirror.

However, there are additionally three undesired terms (B-D). The second term (B) arises from interference of the two reflections from the mirror and produces a large coherent background. The third term (C) results from two reflections from the object itself, and the last term (D) is the conjugate of (A) and will result in a defocused coherent background in the reconstruction. Without coherence-gating it would be very challenging to impossible to filter out the undesired interference terms. However, the four interferences occur at four distinct time delays between the reference arm and the object illumination arm, $\tau_A = \frac{2z_0}{c}, \tau_B = 0, \tau_C = 0, \tau_D = -\frac{2z_0}{c}$.

Using a source with a coherence length of $l_c \ll 2z_0$ allows to effectively suppress the unwanted interference terms by setting the time-delay between the two arms to $\tau = \tau_A$. A shorter coherence length source would be beneficial to both improve the axial-sectioning resolution and reduce the minimal working distance.

**Digital filtering of MCF pixelation**

Due to the low fill-factor of the MCF cores, the holographically reconstructed field is strongly pixelated at the fiber distal facet plane (Fig. 1 B, inset). This pixelation manifest itself as "ghosts" replicas in the images reconstructed by backpropagating the pixelated raw fields (Fig. S2 C), which are the result of the MCF under-sampling the fields. These are digitally filtered in a straightforward manner by applying a low-pass filter to the 2D Fourier transform of the reconstructed fields, i.e., by Fourier-interpolating the holographically measured fields, effectively interpolating the fields between the cores at the fiber facet (Fig. S2 D). The cutoff frequency of the low-pass filter is set to $k_{cutoff} = \pi/p$ where $p$ is the core-to-core pitch. The Fourier interpolation also effectively limits the detection NA to $\sim \lambda/2p$, which dictates the reconstructed fields resolution.

The presented results were obtained using an MCF with identical cores positioned on an ordered grid with very low crosstalk (Schott 153385). Similar results were obtained an MCF with inhomogeneous cores positioned on an imperfectly ordered grid (Fujikura FIGH-06-300S) (Fig.

S4). In this case of imperfectly ordered grid, the spatial cutoff frequency of the Fourier interpolation filter was set according to the average pixel pitch. A discussion on the effects of the various fiber parameters is presented in the Supplementary materials.

## SUPPLEMENTARY MATERIALS

**Section S1**: Theoretical resolution and field-of-view analysis.
**Section S2**: Choice of optimal system parameters
**Fig S1**: Experimental setup
**Fig S2**: Fourier Interpolation on ordered MCF
**Fig S3**: Resolution and field-of-view characterization, detailed figure
**Fig S4**: Imaging using a disordered MCF
**Movie S1**: Imaging a moving USAF target at 50 FPS.
**Movie S2**: Imaging a moving USAF target while bending the MCF.

## ACKNOWLEDGMENTS

**General**: We thank the Hebrew University Center for Nanoscience and Nanotechnology for distal mirror fabrication, and Tali Brooks for proofreading the manuscript.

**Funding:** This work received funding from the European Research Council (ERC) Horizon 2020 research and innovation program (grant no. 677909), Azrieli foundation, Israel Science Foundation (1361/18), Israeli Ministry of Science and Technology.

**Author contributions:** NB and OK conceived the idea and performed analytical analysis, NB performed numerical simulations, experimental measurements, and data analysis. NB and OK wrote the manuscript.

**Competing interests:** The authors declare no competing interests.

**Data and materials availability:** All data needed to evaluate the conclusions in the paper are present in the paper and/or the Supplementary Materials.

# SUPPLEMENTARY MATERIALS
## Section S1: Theoretical resolution and field-of-view analysis

**Resolution**

Collecting the intensity through MCF limits the system resolution due to both the diameter of the fiber, D, and the maximum accepted numerical aperture $NA_{max}$. Each core NA will limit the maximal NA of the collected light to a maximum NA of $NA_{max}$ for objects closer then $\frac{D}{2NA_{max}} - z_m$, i.e. $NA_{eff} = NA_{max}$. For objects at larger distances, the NA will be limited by the bundle diameter to $NA_{eff} = \frac{D}{2(z_o+z_m)}$. The theoretical diffraction-limited resolution of the system can thus be summed up as follows:

$$\sigma_x = \frac{\lambda}{2NA_{eff}} = \begin{cases} \lambda \frac{1}{2NA_{max}} & z_0 < \frac{D}{2NA_{max}} - z_m \\ \lambda \frac{(z_m + z_o)}{D} & otherwise \end{cases} \quad (S1)$$

To validate the theoretical estimation and evaluate the experimental system resolution, we performed a knife-edge measurement, where a 'step-like' reflective target with a sharp edge was imaged (Fig. S3 A). The Point Spread Function (PSF) was calculated from the spatial derivative of the reconstructed target image for each image row, as shown in Fig. S3 A, B. The results for the obtained resolution as a function of the target distance, shown in Fig. 3J, are the FWHM of the calculated PSF averaged over the image rows, and are in good agreement with the theoretical prediction, given in Eq.S1. The error-bars of Fig. 3J are the standard deviation of the PSF FWHM, calculated at the different rows.

**Field of View**

The theoretical Field of view ($FOV$) limitations of the system can be estimated using geometrical considerations: First, the object must be illuminated by the illumination field, which has a NA of the first mode of the MCF core ($NA_1$), i.e. $FOV \approx 2NA_1(z_o + z_m)$. For diffuse reflecting objects, this is the only limitation, and the spatial field of view can theoretically be larger than the MCF diameter, $D$. However, for specular targets that are parallel to the fiber facet, the reflection angle is similar to the incidence angle of the illumination light (neglecting the effects of diffraction), and light will be collected by the MCF only from reflectors that are located at a $FOV < \frac{D}{2}$. To sum up the theoretical considerations predict a FOV that is:

$$FOV = \begin{cases} \frac{D}{2} & a\ specular\ target \\ 2NA_1(z_o + z_m) & a\ diffuse\ target \end{cases} \qquad (S2)$$

As explained in the next section, in our experiments the mirror distance, $z_m$, is chosen such that $4NA_1 z_m \geq D$, which means that $FOV \geq D/2$ for both specular and diffusive targets.

To validate the theoretical estimation and characterize the experimental system FOV, we characterized the FOV of the experimental system using an extended mirror as the imaged object. The mirror was reconstructed using digital back-propagation, but without normalizing by $|E_{illum}|$, leaving a Gaussian-like intensity profile, which its width can be used as a measure for the system FOV. As a figure of merit we calculated the $1/e^2$-width of the reconstructed intensity profile. The result of FOV measurements performed at multiple distances, shown in Fig.3I show an almost constant FOV at all distances, as theoretically expected. The practical FOV can be larger than this estimate since the exact definition of the FOV dictated by the required Signal to Noise Ratio (SNR), which is dependent on the illumination intensity and target reflectivity. In our experiments for imaging a reflecting USAF target using a $D = 650\mu m$ MCF, details of the target are clear at a FoV exceeding $300\mu m$ at all distances as shown in Fig. S3 C.

Each data point in Fig. 3I is the average $1/e^2$-width of the reconstructed image of the mirror, averaged at 6 different angles in a single reconstructed image, at intervals of 30 degrees. The error-bars in Fig. 3I are the standard deviation of the $1/e^2$-width at these 6 angles.

## Section S2: Choice of optimal system parameters

### Distal mirror distance

To choose the mirror distance, $z_m$, both the phase variation and the intensity profile of $E_{ref}$ must be considered. First, the phase of the field is a spherical one, with a curvature of $2z_m$: $\phi(E_{ref}) = \frac{2\pi}{\lambda 4 z_m}(x^2 + y^2)$. To ensure a well-defined unvarying reference phase at each core, we demand that $d\frac{\partial}{\partial x}\phi(E_{ref}) \ll \pi$ over the entire distal facet. Differentiating the above expression for $\phi(E_{ref})$, and plugging $x_{max} = D/2$ results in the following demand for the mirror distance:

$$z_m \gg \frac{Dd}{2\lambda} \quad (S3)$$

In addition, the intensity profile must be wide enough to cover the entire fiber, i.e., requiring that $z_m \geq \frac{D}{4NA_1}$. The first requirement (eq. S3) automatically fulfills the latter, since $NA_1 \leq \frac{\lambda}{2d}$.

### Distal mirror reflectively and reference- and signal-arms powers ratio

Three important system parameters affect the SNR of the measurements and can be determined by the demand to maximize the SNR of the system. First, the reflectively of the partially reflecting distal mirror, $R$, i.e., the power reflection coefficient. Second, the total illumination power emerging from the distal facet, $P$. The third parameter, A, is the fraction of the total illumination power which is in the reference arm. For the sake of simplicity, no absorption is considered, and an object with a total effective power reflection coefficient of $R_o \ll 1$ is considered. Where $R_o$ also takes into account the power lost to propagation and effective reflective area of the object. For further ease of notations, we denote the power ratio in the object arm $B = 1 - A$, and the power transmission coefficient of the mirror $T = 1 - R$.

To analyze the optimal SNR, one may consider the signal collected by a single core. In each single core, four fields are superimposed, as described in "Suppression of spurious reflections" section. Using the same notations, we analyze the expected amplitude of each of them:

$$\begin{aligned} |E_{R,m}| &= \sqrt{PA}\sqrt{R} \\ |E_{R,o}| &= \sqrt{PA}\sqrt{T^2 R_o} \\ |E_{I,m}| &= \sqrt{PB}\sqrt{R} \\ |E_{I,o}| &= \sqrt{PB}\sqrt{T^2 R_o} \end{aligned} \quad (S4)$$

Where $E_{R,m}$ and $E_{R,o}$ are the reflections of the reference beam from the mirror and object, respectively. $E_{I,m}$ and $E_{I,o}$ are the reflections of the illumination beam from the mirror and object, respectively.

The desired signal is the interference term between $E_{R,m}$ and $E_{I,o}$, while the constant background is proportional to all four intensities. Assuming the dominant source of noise in the camera detected signal is shot-noise from the strong background, as is commonly the case when using sCMOS cameras, the SNR in a single frame will be:

$$SNR \propto \frac{|E_{R,m}E_{I,o}|}{\sqrt{E_{R,m}^2 + E_{R,o}^2 + E_{I,m}^2 + E_{I,o}^2}} = \sqrt{P} \cdot \sqrt{AB} \cdot \frac{\sqrt{RT^2 R_o}}{\sqrt{R + T^2 R_o}} \qquad (S5)$$

The SNR is thus a product of three terms: the first is the total illumination power, that is limited in most practical applications by sample damage. The second term $\sqrt{AB} = \sqrt{A(1-A)}$ is maximized by setting $A = B = 0.5$, i.e., setting the powers in the two arms to be equal with a balanced MZI. Lastly, the third term in eq. S5 is maximized by setting R and T to fulfill:

$$\frac{R^2}{T^3} = \frac{R_o}{2} \qquad (S6)$$
$$\Rightarrow R \approx \sqrt{R_o/2}$$

The last approximation is valid only assuming $R \ll 1$, which is indeed the case when considering weakly-reflecting objects. For example, for $R_o = 0.02$, equation S6 yields $R \approx 0.1$. In our experiments $R$ is determined in the mirror fabrication process. For the results shown throughout this article, $R \approx 0.12$ was achieved by evaporating $10_{nm}$ of titanium on a microscope slide. To fine-tune the power ratio of the two arms, $A$, a polarizing beam-splitter along with two half-wave-plates was incorporated into the setup, as explained in the "Experimental design" section.

**Fiber parameters**

Commercial MCFs are characterized by several important parameters: the pitch, $p$, the single core NA, the number of modes in a single core, the total fiber diameter, $D$, and the intracore crosstalk. To use FiDHo, the correct fiber and mirror distance must be chosen. First, to maximize resolution, a large $NA_{max}$ is needed, as achieved with larger core sizes. In addition, increasing the bundle size $D$ will increase both $NA_{eff}$ at larger angles and the $FOV$ for weakly scattering targets, at the price of a larger footprint. Smaller pitch will improve resolution, since less spatial interpolation will be needed on the fiber facet. Lastly, to reduce bending sensitivity and to increase resolution, a fiber with minimal crosstalk should be chosen, while maintaining a reasonable fill factor. For the fibers and laser used throughout this article FIGH-06-300S, $D = 270 \mu m$, $NA_{core} \sim 0.3$, and a subset of Schott 1533385, $D = 650 \mu m$, $NA_{core} \sim 0.3$, $\lambda = 640_{nm}$. In all experiments, a mirror distance of $z_m = 2000 \mu m$ was used.

**Fig. S1**

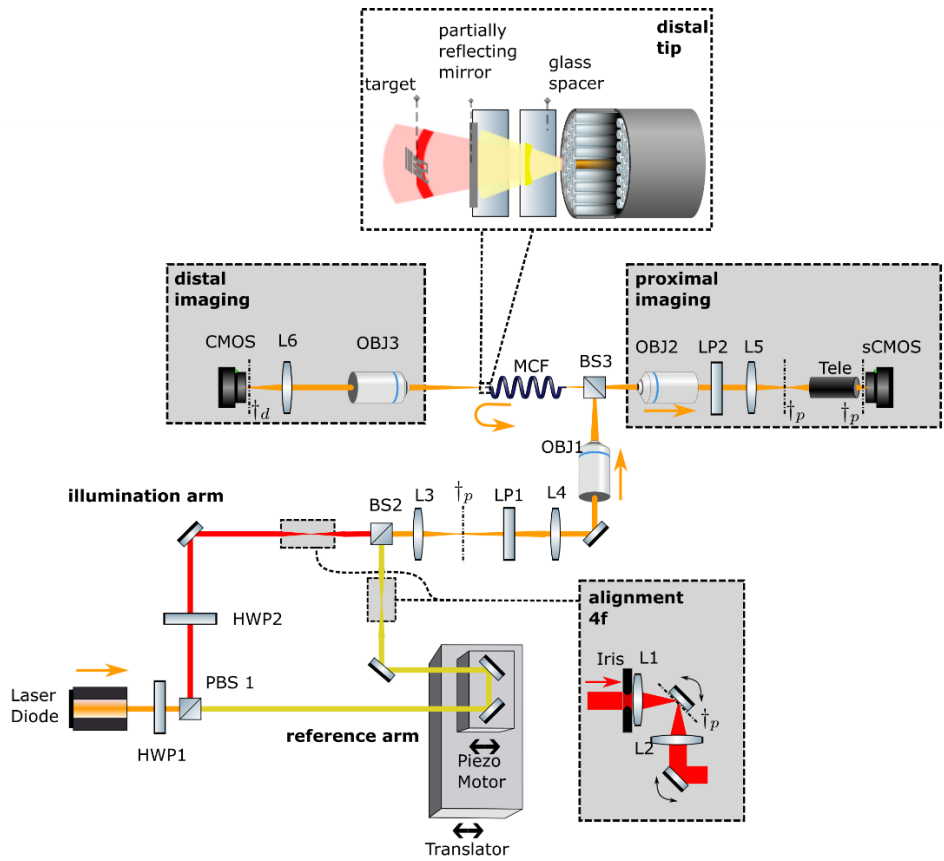

**Fig. S1. Experimental setup.** See section " Experimental Design " for complete description. HWP half-wave plate, PBS polarizing beam-splitter, BS beam-splitter, L lens, OBJ objective, MCF multicore fiber, Tele telescope, $\dagger_p$, $\dagger_d$ mark the proximal and distal conjugate planes, respectively.

*s*

**Fig. S2**

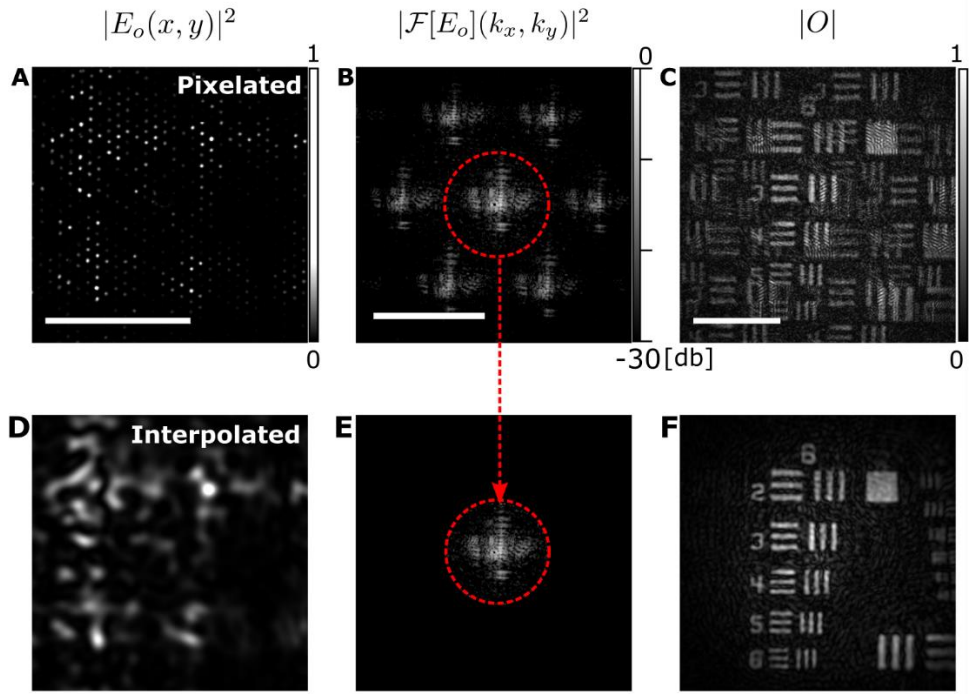

**Fig. S2. Fourier Interpolation on ordered MCF** (**A**) The retrieved distal field on a Schott fiber used throughout the article, shows the individual cores with a constant inter-core pitch. (**B**) In the frequency domain, clear replicas appear due to ordered sampling of the field. The red circle marks the frequencies to be saved. (**C**) Reconstruction of the object with no interpolation, shows ghosting of the object, due to aliasing. (**D, E, F**) show the same data, after filtering out higher spatial frequencies. Scale bars: A, C - $100_{\mu m}$  B - $\frac{2\pi}{pitch}$

**Fig. S3**

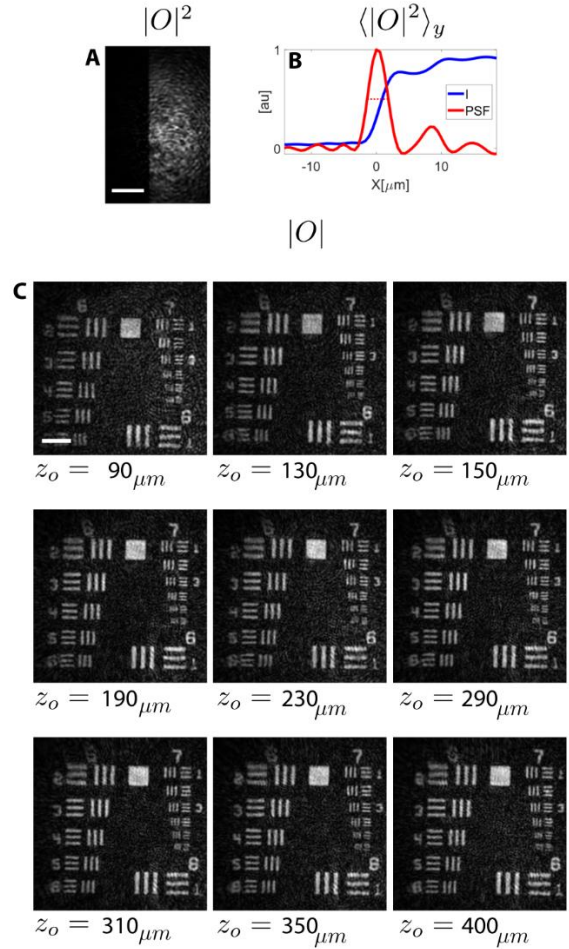

**Fig. S3. Resolution and field-of-view characterization, detailed figure** (**A**) a single image of the knife-edge mirror, using the edge of a USAF group 3 square, at a distance of $z_0 = 280 \mu m$ (**B**) The mean cross section of (**A**)(blue), and the PSF (red) calculated as the spatial derivative of the cross section (red). The FWHM of the PSF, is indicated by the dashed line. (**C**) Imaging a USAF target at different depths, demonstrate the constant resolution and field of view. Scale bars: $50 \mu m$

**Fig. S4**

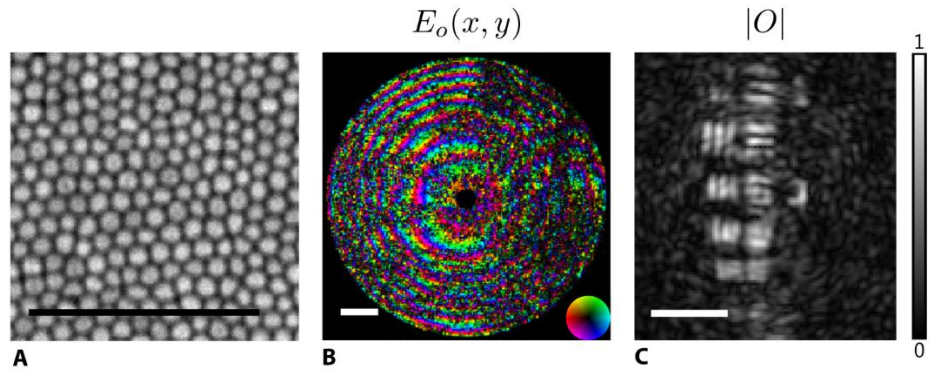

**Fig. S4. Imaging using a disordered MCF** (**A**) White light image of the FIGH-06-300S Fujikura fiber used (**B**) The retrieved distal field on the same fiber, when placing a reflective USAF resolution target at a distance of $z_0 = 100 \mu m$ with the mirror placed at $z_m = 2_{mm}$. The cores are indiscernible due to binning of the camera pixels (**C**) The reconstructed object amplitude. Scale bars $40 \mu m$.